\documentclass[twocolumn,prb,aps,psfig,showpacs,preprintnumbers,superscriptaddress]{revtex4}
\usepackage{graphicx}
\usepackage{epstopdf}
\usepackage{float}
\usepackage{color}
\usepackage{amsmath}

\begin{document}

\title{Coexistence of antiferromagnetic and ferromagnetic spin correlations in SrCo$_2$As$_2$ revealed by $^{59}$Co and $^{75}$As NMR}

\author{P.~Wiecki}
\affiliation{The Ames Laboratory, Ames, IA 50011, USA}
\affiliation{Department of Physics and Astronomy, Iowa State University, Ames, IA 50011, USA}

\author{V.~Ogloblichev }
\affiliation{The Ames Laboratory, Ames, IA 50011, USA}
\affiliation{Institute of Metal Physics, Ural Division of Russian Academy of Sciences, Ekaterinburg 620990, Russia}

\author{Abhishek Pandey}
\affiliation{The Ames Laboratory, Ames, IA 50011, USA}
\affiliation{Department of Physics and Astronomy, Iowa State University, Ames, IA 50011, USA}

\author{D.~C.~Johnston}
\affiliation{The Ames Laboratory, Ames, IA 50011, USA}
\affiliation{Department of Physics and Astronomy, Iowa State University, Ames, IA 50011, USA}

\author{Y.~Furukawa}
\affiliation{The Ames Laboratory, Ames, IA 50011, USA}
\affiliation{Department of Physics and Astronomy, Iowa State University, Ames, IA 50011, USA}

\date{\today}

\begin{abstract}

      In nonsuperconducting, metallic paramagnetic SrCo$_2$As$_2$, inelastic neutron scattering measurements have revealed strong stripe-type $\mathbf{q} = (\pi,0)$ antiferromagnetic (AFM) spin correlations. 
      Here, using nuclear magnetic resonance (NMR) measurements on $^{59}$Co and $^{75}$As nuclei, we demonstrate that stronger ferromagnetic (FM) spin correlations coexist in SrCo$_2$As$_2$. 
     Our NMR data are consistent with density functional theory  (DFT) calculations which show enhancements at both $\mathbf{q} = (\pi,0)$  and the in-plane FM $\mathbf{q} = 0$ wavevectors in static magnetic susceptibility $\chi(\bf {q})$.
      We suggest that the strong FM fluctuations prevent superconductivity in SrCo$_2$As$_2$, despite the presence of stripe-type AFM fluctuations. 
    Furthermore, since DFT calculations have consistently revealed similar enhancements of the $\chi(\bf {q})$ at both $\mathbf{q} = (\pi,0)$ and $\mathbf{q} = 0$ in the iron-based superconductors and parent compounds, our observation of FM correlations in SrCo$_2$As$_2$ calls for  detailed studies of FM correlations in the iron-based superconductors.

\end{abstract}

\pacs{74.70.Xa, 76.60.-k, 75.40.Gb}
\maketitle
    The interplay between magnetism and superconductivity is one of the central issues in unconventional superconductors (SCs) such as high $T_{\rm c}$ cuprates and iron pnictide-based SCs.
     Among the iron pnictide-based SCs, the ``122'' class of compounds, \textit{A}Fe$_2$As$_2$ (\textit{A} = Ca, Ba, Sr, Eu), has been one of the most  widely studied systems in recent years. \cite{Johnston2010,Ni2008,Rotter2008,Sefat2008,Canfield2010,Stewart2011,Ni2008_2}
    These systems undergo coupled structural and magnetic phase transitions at a system-dependent N\'eel temperature $T_{\rm N}$, below which long-range stripe-type antiferromagnetic (AFM) order emerges. 
    Suppression of the AFM order by doping or pressure results in a SC ground state with  $T_{\rm c}$ ranging from a few K to more than 50~K. 
    Continued doping ultimately results in the suppression of the stripe-type AFM spin fluctuations, which correlates with the suppression of SC.\cite{Johnston2010,Ni2008,Rotter2008,Sefat2008,Canfield2010,Stewart2011,Ni2008_2}
     Although the Cooper pairing is widely believed to originate from the residual stripe-type AFM spin fluctuations, the origin of the large variability of $T_{\rm c}$ is still not well understood.

    Tetragonal, metallic paramagnetic (PM) SrCo$_2$As$_2$ is the end member of the electron-doped Sr(Fe$_{1-x}$Co$_x$)$_2$As$_2$ family of compounds, which displays superconductivity in the range from $x = 0.07$ to $x = 0.17$ with a maximum $T_{\rm c}$ of $19$ K.\cite{Jasper2008,Hu2011}
    The $x=0$ parent compound, SrFe$_2$As$_2$, is an AFM showing stripe-type spin density wave order below 220 K.\cite{Krellner2008,Zhao2008} 
    In SrCo$_2$As$_2$, on the other hand, no long range magnetic ordering is observed down to 1.8 K.\cite{pandey} 
   The Sommerfeld coefficient ($\gamma=37.8$  $\frac{\rm mJ}{\rm mol \cdot K^2}$) is significantly
enhanced relative to SrFe$_2$As$_2$ in the stripe AFM state.\cite{pandey}
    Angle-resolved photoemission spectroscopy  and electronic structure calculations show no clear nesting features of the Fermi surface which drive the stripe-type AFM order and SC in the parent and modestly doped compounds.\cite{pandey}
     Nevertheless, AFM spin correlations  are suggested from the temperature ($T$) dependence of  magnetic susceptibility $\chi$  which exhibits a broad maximum around 115 K, a characteristic of short-range dynamic AFM correlations in low-dimensional spin systems. \cite{pandey}
    Subsequent inelastic neutron scattering (INS) measurements on SrCo$_2$As$_2$ revealed strong AFM spin fluctuations at the stripe-type wavevector.\cite{INS}
    Similar physical properties are reported in the SC compound KFe$_2$As$_2$, the end member of the hole-doped Ba$_{1-x}$K$_x$Fe$_2$As$_2$ family. 
   This compound also has an enhanced $\gamma=103$ $\frac{\rm mJ}{\rm mol \cdot K^2}$ 
    and a broad peak in $\chi$ around $100$ K,\cite{Hardy2013} along with strong stripe-type AFM fluctuations.\cite{Hirano2012} 
    The similarity between SC KFe$_2$As$_2$ and non-SC SrCo$_2$As$_2$ raises the important question of why superconductivity does not arise in SrCo$_2$As$_2$.

      In this paper, we report $^{59}$Co and $^{75}$As nuclear magnetic resonance (NMR) measurements to examine the local microscopic properties of SrCo$_2$As$_2$. 
      Our analysis, based on the modified Korringa relation, reveals strong ferromagnetic (FM) spin fluctuations within the Co layer 
      coexisting with the stripe-type AFM fluctuations observed by INS. 
      Based on these results, we suggest that the low-energy FM fluctuations observed by NMR compete with the stripe-type AFM fluctuations, resulting in the suppression of SC in SrCo$_2$As$_2$. 
      Furthermore, our observation of coexisting stripe AFM and FM fluctuations in SrCo$_2$As$_2$ is consistent with density functional theory (DFT) calculations which show peaks in the static susceptibility, $\chi(\mathbf{q})$, at both the FM and stripe AFM in-plane wavevectors.\cite{INS} 
    This theoretically predicted enhancement of $\chi(\mathbf{q})$ at both wavevectors is not unique to SrCo$_2$As$_2$ but in fact applies more generally to iron-pnictide based superconductors and parent  compounds.\cite{Johnston2010,Singh2008,Mazin2008,Dong2008,Yaresko2009,Neupane2011}
      Our NMR data provide the first microscopic confirmation of spin susceptibility enhanced at both wavevectors in the iron-pnictide family, indicating that FM fluctuations may play an important role in determining $T_{\rm c}$ in iron-pnictide based SCs.

     NMR measurements were performed on $^{75}$As ($I=3/2$, $\gamma/2\pi=7.2919$ MHz/T) 
and $^{59}$Co ($I=7/2$, $\gamma/2\pi=10.03$ MHz/T) using a homemade phase-coherent spin-echo pulse spectrometer.
     The $^{59}$Co and $^{75}$As spin-lattice relaxation rates ($1/T_1$) were measured with a recovery method using a single $\pi$/2 saturation pulse.\cite{T1} 
     The single crystal used in this study was grown with Sn flux and is same as that in our previous study \cite{pandey} where preliminary $^{75}$As-NMR results were reported.  

\begin{figure}[t]
\centering
\includegraphics[width=8.5cm]{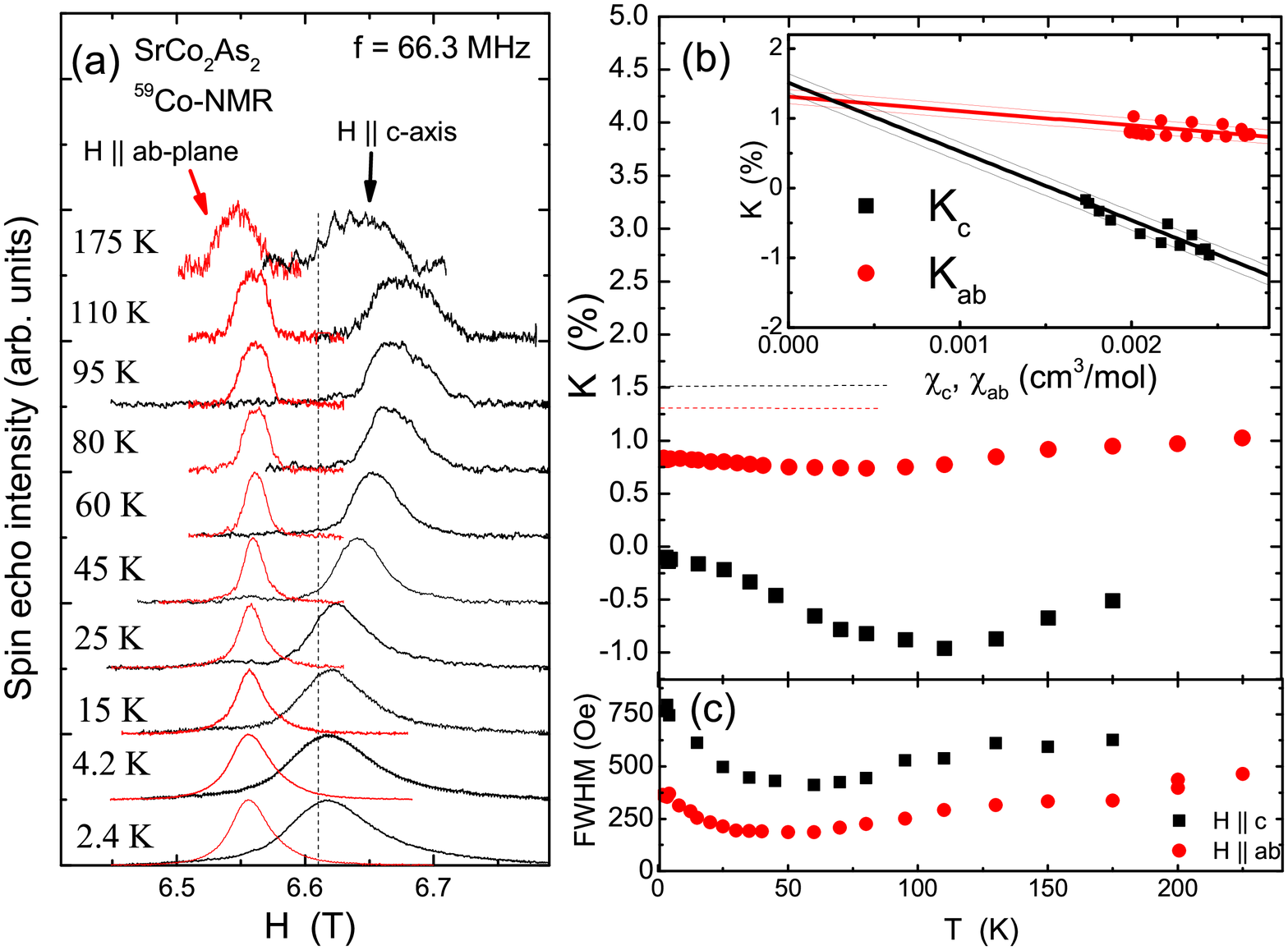}
\caption{(Color online) (a) Field-swept $^{59}$Co-NMR spectra at frequency $f=66.3$ MHz for magnetic fields 
$H\parallel c$ axis (black) and $H\parallel ab$ plane (red)  to at various values of $T$. The vertical dashed line represents the zero-shift position ($K=0$).
(b) $T$ dependence of the $^{59}$Co-NMR shifts $K_c$ and $K_{ab}$. The black and red dashed lines are corresponding to $K_0$ for $K_c$ and $K_{ab}$, respectively. Inset: $K$ vs $\chi$ plots for each field direction where we used $\chi$ data reported in Ref. [\onlinecite{pandey}]. The thick solid lines are fitting results and two thin lines above and below the thick line give an error for our estimate of $K_0$ for each $H$ direction. (c) $T$ dependence of the full width at half maximum (FWHM) of the spectra for each field direction. }
\label{fig:spectrum}
\end{figure}

      Figure \ref{fig:spectrum}(a) shows field-swept $^{59}$Co-NMR spectra at various values of $T$ for magnetic fields parallel to the $c$~axis ($H\parallel c$~axis) and to the $ab$~plane ($H\parallel ab$~plane).  
    The typical spectrum for a nucleus with spin $I=7/2$ with Zeeman and quadrupolar interactions can be described by a nuclear spin Hamiltonian ${\cal{H}}=-\gamma\hbar\mathbf{I}\cdot\mathbf{H}_{\text{eff}}+\tfrac{h\nu_Q}{6}[3I_z^2-I(I+1)]$,
where $\mathbf{H}_{\text{eff}}$ is the effective field at the nuclear site and $h$ is Planck's constant. 
   The nuclear  quadrupole frequency for $I=7/2$ nuclei is given by $\nu_{\rm Q} = e^2QV_{\rm ZZ}/14h$, where $Q$ is the nuclear quadrupole moment and $V_{\rm ZZ}$ is the electric field gradient at the nuclear site.
    For $I=7/2$ nuclei, this Hamiltonian produces a spectrum with a sharp central transition line flanked by three satellite peaks on either side. 
   The observed $^{59}$Co NMR spectra, however, do not show the seven distinct lines but rather exhibit a single broad line due to inhomogeneous broadenings.
    From the line width, we estimate $\nu_{\rm Q}$ $\sim$ 0.14 MHz at 4.2 K with $V_{\rm ZZ}$ parallel to the $c$~axis,  
    close to the value of 0.13 MHz for $^{59}$Co in Ba(Fe$_{1-x}$Co$_x$)$_2$As$_2$ with $x$ = 0.02 and 0.04.\cite{Ning2009}   

\begin{figure}[t]
\centering
\includegraphics[width=7.5cm]{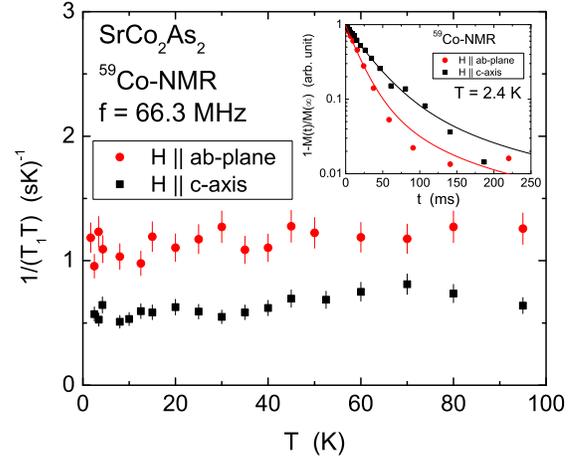}
\caption{(Color online) $T$ dependence of $1/T_1T$ for both magnetic field directions, $H\parallel c$ axis [1/($T_1T)_{H\|c}$] and  $H\parallel ab$ plane [1/($T_1T)_{H\|ab}$].
Inset: Recovery curves at $T=2.4$ K for both $H$ directions. The solid lines are fits by the relaxation function described in the text.}
\label{fig:T1T}
\end{figure}

    Figure \ref{fig:spectrum}(b) shows the $T$ dependence of the NMR shift for $H\parallel c$~axis ($K_c$) and  $H\parallel ab$ plane ($K_{ab}$). 
    The NMR shift has contributions from the $T$-dependent spin part $K_\text{spin}$ and a $T$-independent orbital part $K_0$. 
     $K_\text{spin}$ is proportional to the spin susceptibility $\chi_{\rm spin}$ through the hyperfine coupling constant $A_{\rm{hf}}$ giving $K(T)=K_0+\frac{A_{\rm {hf}}}{N_{\rm A}}\chi_{\text{spin}}(T)$, where $N_{\rm A}$ is Avogadro's number.
    The anisotropic spin susceptibilities $\chi_{ab}$ and $\chi_{c}$ in SrCo$_2$As$_2$ were reported in Ref. \onlinecite{pandey}.
     The inset of Fig. \ref{fig:spectrum}(b) plots $K_{ab}$ and $K_{c}$ against $\chi_{ab}$ and $\chi_{c}$, respectively, with
$T$ as an implicit parameter. $T$ is chosen to be above 20 K to avoid upturns in $\chi$ due to impurities.\cite{pandey}
    $K_{ab}$ and $K_c$ vary with the corresponding $\chi$ as expected, although one can see a slight deviation from the linear relationship. 
    We estimated the hyperfine coupling constants $A_{c}= (-110\pm 5) $ kOe/$\mu_{\rm B}$ and $A_{ab}= (-22.9 \pm 1.0) $ kOe/$\mu_{\rm B}$ by fitting the data (shown by the thick lines in the inset).
     $A_c$ is comparable to the value of $-105 $ kOe/$\mu_{\rm B}$ for isotropic $d$ electron core polarization, while $A_{ab}$ is much smaller.    
     The small value of $A_{ab}$ could be due to anisotropic and positive orbital and/or dipolar hyperfine coupling contributions which cancel a part of the negative core polarization hyperfine field.
     Similar reductions in the hyperfine coupling constant have been observed in several Co compounds, which have been well explained by taking the orbital contributions into consideration.\cite{Tsuda1968,Fukai1996,Roy2013}
     The orbital shifts  $K_{0,ab}=(1.31 \pm 0.10 )\%$ and $K_{0,c}=(1.51 \pm 0.13) \%$ were estimated from the fittings.
     In order to estimate the error in $K_0$, we change the $K_0$ while keeping the same slope to cover all data points. 
     The two thin lines correspond to the fitting lines with minimum and maximum $K_0$ for each $H$ direction.   
     The $T$ dependences of $K_{ab}$ and $K_{c}$ are similar to the behaviors reported for $^{75}$As-NMR in Ref. [\onlinecite{pandey}], which show broad maxima at $T \sim 115$~K.
    These maxima are observed as minima in the $^{59}$Co NMR shift data due to the negative hyperfine coupling constant. 
    The broad minima in $K_{ab}$ and $K_{c}$ suggest the presence of low-dimensional dynamic short-range AFM correlations below $115$ K. 
    In Fig. \ref{fig:spectrum}(c), we plot the full width at half maximum (FWHM) of the spectra as a function of $T$ for 
$H\parallel c$ axis and  $H\parallel ab$ plane. 
    With decreasing $T$, the FWHM decreases gradually and starts to increase below  $\sim 30$ K where $\chi$ shows $T$-independent behavior, suggesting inhomogeneities of the dynamic short-range AFM order below $30$ K. 


    To investigate the dynamical magnetic properties, we have measured 1/$T_1$ versus $T$ (Fig.~\ref{fig:T1T}). 
    In both field directions $1/T_1T$ is roughly constant over the entire temperature range. 
    The inset shows nuclear magnetization recovery curves for the two magnetic field directions together with fitting results.   
   To examine the character of the spin fluctuations in more detail, we perform a modified Korringa relation analysis. 
Within a Fermi liquid picture, $1/T_1T$ is proportional to the square of the density of states at the Fermi energy ${\cal D}(E_{\rm F})$ and $K_{\text{spin}} (\propto \chi_{\text{spin}}$) is proportional to ${\cal D}(E_{\rm F})$. 
    In particular, $T_1TK_{\text{spin}}^2$  = $\frac{\hbar}{4\pi k_{\rm B}} \left(\frac{\gamma_{\rm e}}{\gamma_{\rm N}}\right)^2$ = $S$, which is the Korringa relation.  
    Deviations from $S$ can reveal information about electron correlations in the material \cite{Moriya1963,Narath1968}, which are expressed via the parameter $\alpha=S/(T_1TK_{\text{spin}}^2)$. 
      For instance, enhancement of $\chi(\mathbf{q}\neq 0)$ increases $1/T_1T$ but has little or no effect on $K_{\text{spin}}$, which probes only the uniform $\chi$ with $\mathbf{q}$ = 0.  
    Thus  $\alpha >1$ for AFM correlations and $\alpha <1$ for FM correlations.

\begin{figure}[t]
\centering
\includegraphics[width=8cm]{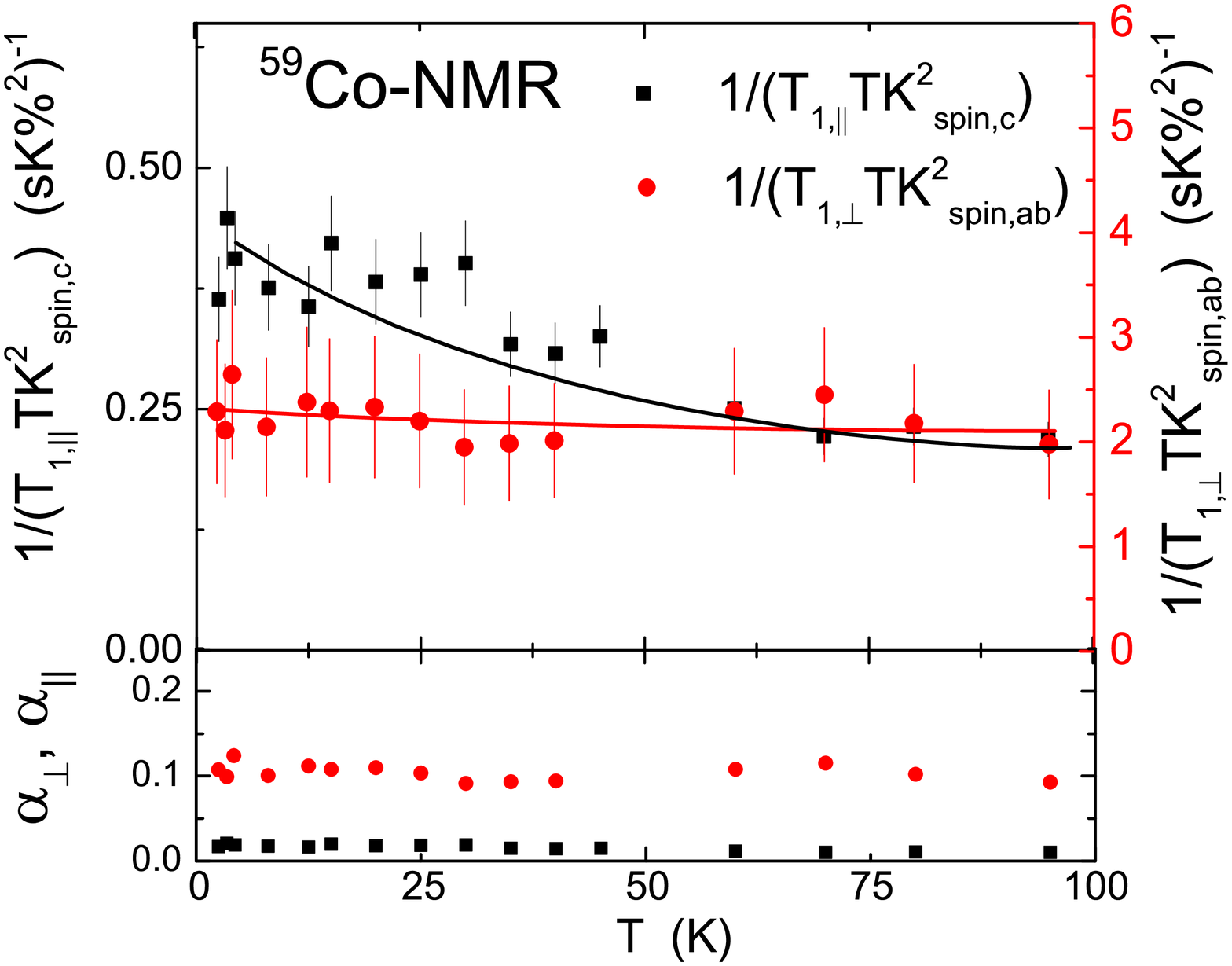}
\caption{(Color online) Top panel: $T$ dependence of the Korringa ratios $1/T_{1,\bot}TK_{\text{spin},ab}^2$ (red) and
$1/T_{1,\|}TK_{\text{spin},c}^2$ (black) for spin correlations in the $ab$ plane and along the $c$ axis, respectively.
   The solid lines are guides to the eye.
Lower panel: $T$ dependence of the parameter $\alpha_{\bot}$ for spin correlations in the $ab$ plane (red) and $\alpha_{\|}$ along the $c$ axis (black). }
\label{fig:T1TK2}
\end{figure}

    Application of the Korringa relation to SrCo$_2$As$_2$ requires some care due to the anisotropy of $K_\text{spin}$ and $1/T_1T$.
    Since $1/T_{1}T$ probes magnetic fluctuations perpendicular to the magnetic field,\cite{Moriya1963_2} it is natural to consider the Korringa ratio $1/T_{1,\bot}TK_{\text{spin},ab}^2$ where $1/T_{1,\bot}T$ = $1/(T_{1}T)_{H\|c}$,  when examining the character of magnetic fluctuations in the $ab$ plane. 
   Similarly, we consider the Korringa ratio $1/T_{1,\|}TK_{\text{spin},c}^2$ for magnetic fluctuations along the $c$ axis. 
   Here $1/(T_{1,\|}T)$ is estimated from  $2/(T_{1}T)_{H\|ab}$ $-$ $1/(T_{1}T)_{H\|c}$.

    In the top panel of Fig. \ref{fig:T1TK2}, we show the $T$ dependence of the Korringa ratios for magnetic fluctuations in the $ab$ 
plane and along the $c$ axis, along with the corresponding values of the parameter $\alpha$ in the bottom panel. 
    We find that $\alpha \ll 1$ in each case, with the value of $\alpha$ remaining constant throughout the range of $T$. 
    The low values of $\alpha$ indicate that the fluctuations have predominantly FM character.  
    It should be emphasized  that the $\alpha$ values strongly depend on $K_{\rm spin}$ and the $\alpha_{\bot}$  greater than $\alpha_{\|}$ could be due to the small $K_{\rm spin}$ values arising from the small $A_{ab}$. 
    In addition, it should be noted that the observed 1/$T_1T$ is the sum of four contributions: the $s$ electron Fermi contact, $d$ orbital, $d$ core polarization, and $d$ dipole relaxation rates.     
    As a result,  the estimated values for $\alpha$ for both directions can be considered to be upper limits on $\alpha$, indicating even stronger FM fluctuations in SrCo$_2$As$_2$ than expected from the above $\alpha$ values. 
    On the other hand, the increase of $1/T_{1,\|}TK_{\text{spin},c}^2$ below $50$ K clearly indicates the presence of AFM correlations along the $c$~axis coexisting with the dominant FM fluctuations.


\begin{figure}[t]
\centering
\includegraphics[width=6.5cm]{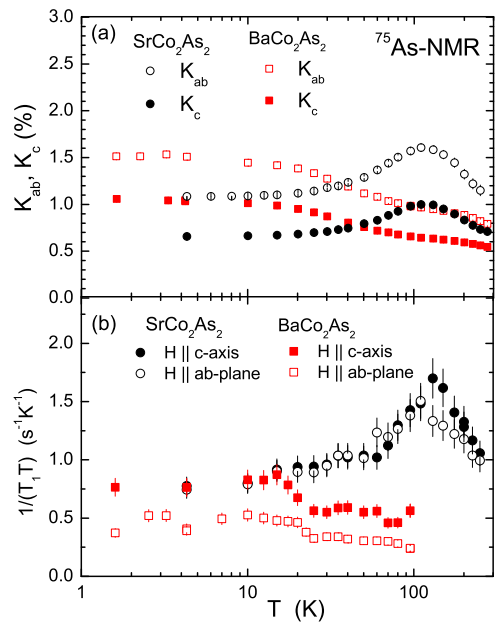}
\caption{(Color online) (a) $T$ dependence of the $^{75}$As NMR shift for both field directions in SrCo$_2$As$_2$ (black)
and  BaCo$_2$As$_2$ (red) \cite{baco2as2yuji}. 
(b) $T$ dependence of $1/T_1T$ at the $^{75}$As sites for both field directions in SrCo$_2$As$_2$ (black) and BaCo$_2$As$_2$ (red).}
\label{fig:75As}
\end{figure}

    The suggested FM spin correlations can be confirmed by $^{75}$As NMR in SrCo$_2$As$_2$.
     In Fig.~\ref{fig:75As}(a), we show the $T$ dependence of the $^{75}$As NMR shifts $K_{ab}$ and $K_{c}$. 
     For comparison, Fig.~\ref{fig:75As} also shows our analogous data from isostructural BaCo$_2$As$_2$ reported in Ref.~\onlinecite{baco2as2yuji}, which are in agreement with those reported in Ref.~\onlinecite{baco2as2imai}.
    The NMR shifts measured at the $^{75}$As sites of SrCo$_2$As$_2$ display broad maxima at $T\sim115$~K, consistent with the NMR shift measured at the $^{59}$Co sites, although with opposite sign of the hyperfine coupling. 
     The broad peak observed in SrCo$_2$As$_2$ contrasts sharply with the NMR shift in BaCo$_2$As$_2$, which increases with decreasing $T$ and then levels off at low $T$. 

\begin{figure}[t]
\centering
\includegraphics[width=7.2cm]{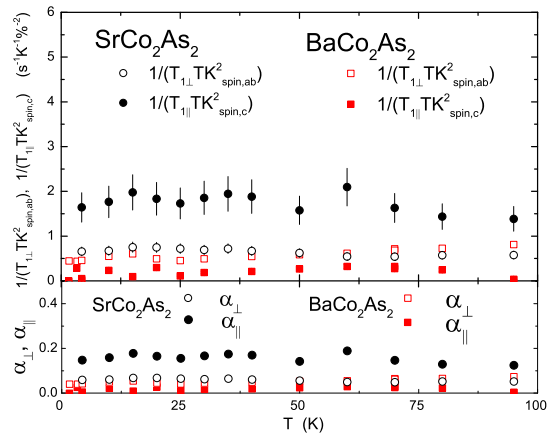}
\caption{(Color online) Top panel: $T$ dependence of the Korringa ratios $1/T_{1,\bot}TK_{\text{spin},ab}^2$ (open symbols) and
$1/T_{1,\|}TK_{\text{spin},c}^2$ (closed symbols) for spin correlations in the $ab$ plane and along the $c$ axis, respectively, for SrCo$_2$As$_2$ (black) and BaCo$_2$As$_2$ (red).
Lower panel: $T$ dependence of $\alpha_{\bot}$ and $\alpha_{\|}$ for SrCo$_2$As$_2$ and BaCo$_2$As$_2$. }
\label{fig:alpha_75As}
\end{figure}

    The $T$ dependence of $1/T_1T$ of $^{75}$As, measured in both field directions, is shown in Fig.~\ref{fig:75As}(b) for both SrCo$_2$As$_2$ and BaCo$_2$As$_2$.
      For SrCo$_2$As$_2$, $1/T_1T$ for both field directions shows a broad peak around $T\sim115$ K. 
     This $T$ dependence is very similar to that of the NMR shift. 
      Also in BaCo$_2$As$_2$, $1/T_1T$ shows a very similar $T$ dependence to that of the corresponding NMR shift. 
     This similar $T$ dependence of $1/T_1T$ and $K$ for BaCo$_2$As$_2$ was also noted in Ref. [\onlinecite{baco2as2imai}]. 
      Ahilan {\it et al.} contrasted this behavior to that of the PM state in optimally-doped Ba(Fe$_{1-x}$Co$_x$)$_2$As$_2$, where $1/T_1T$ increases with decreasing $T$ below 100 K, while $K$ slowly decreases.\cite{baco2as2imai}
      This behavior is clear evidence for the presence of fluctuations with $\mathbf{q}\neq0$. 
    In contrast, the similar $T$ dependence of $1/T_1T$ and $K$ in BaCo$_2$As$_2$ rules out strong fluctuations with $\mathbf{q}\neq0$, since these would increase $1/T_1T$ but not $K$.
      Ahilan {\it et al.} therefore concluded the correlations in BaCo$_2$As$_2$ are primarily FM in nature. 
       By a similar argument, our data on SrCo$_2$As$_2$ offer clear evidence for dominant FM fluctuations.
       In fact, the FM fluctuations can be shown by the modified Korringa relation analysis using the $^{75}$As NMR data. 
    Figure \ref{fig:alpha_75As} shows the Korringa ratios for both field directions in SrCo$_2$As$_2$  and BaCo$_2$As$_2$ along with the corresponding Korringa parameters $\alpha$. 
    In each case we find $\alpha\ll1$, again consistent with strong FM fluctuations in both materials,  consistent with dominant FM correlations as found above for $^{59}$Co in SrCo$_2$As$_2$. 
    The slightly higher value of the $\alpha_{\|}$ for SrCo$_2$As$_2$ than the other three cases suggests that the $c$~axis component of the magnetic fluctuations in SrCo$_2$As$_2$ would be less FM than in BaCo$_2$As$_2$. 
     The above analysis is based on a simple model  that the nuclear relaxation is due to the local ${\cal D}$$(E_F)$ at the As sites, through on-site hyperfine interactions, where As-4$p$ bands hybridize with Fe-3$d$ bands. 
     On the other hand, if the relaxations are induced by only localized Fe electronic spins through isotropic transferred hyperfine interactions, the $\alpha$ value would be modified by a factor of 4 due to the $q$ dependent hyperfine form factor;\cite{Jeglic2010} FM spin correlations would then be expected for $\alpha$ $<$ 0.25.
  Regardless of the model, the $\alpha$ values in both systems are consistent with FM spin correlations.

     According to DFT calculations in Ref.~[\onlinecite{INS}], the $\chi(\mathbf{q})$ in SrCo$_2$As$_2$ shows enhancements of similar strength at both the FM and stripe AFM wavevectors.  
Furthermore, the DFT results indicate that the stripe-type AFM fluctuations have a higher energy scale than the FM fluctuations, suggesting that
FM fluctuations my be dominant at low energies. 
   From the NMR point of view, which probes energies very near the ground state,  we find that the fluctuations are indeed predominantly FM in character. 
     We also find evidence for weak AFM fluctuations coexisting with the dominant FM fluctuations, which can be attributed to the contribution in $\chi(\mathbf{q})$ at the stripe AFM wave vector revealed by the DFT calculations and INS measurements. 

     In summary, our $^{59}$Co and $^{75}$As NMR data demonstrate that the low energy spin fluctuations in paramagnetic SrCo$_2$As$_2$, the end member of the electron-doped Sr(Fe$_{1-x}$Co$_x$)$_2$As$_2$ family,  are predominantly FM in character. 
    We also found clear evidence of coexisting weak stripe-type AFM fluctuations that also appear at the higher INS energy scale. 
    In the standard phenomenology of the iron-arsenide SCs, optimum SC is expected if strong stripe-type AFM fluctuations are present in the absence of long-range AFM order. 
    We suggest that the competing low energy FM fluctuations interfere with the stripe-type AFM fluctuation-based pairing mechanism, thus suppressing superconductivity in SrCo$_2$As$_2$ even though the standard requirements are satisfied. 
    Finally, several theoretical calculations have shown enhancements of $\chi(\mathbf{q})$ at both the FM and stripe-type AFM wavevectors in iron-based superconductors and parent compounds, similar to the case of SrCo$_2$As$_2$.
    Experimentally, a Korringa parameter $\alpha$  from $^{77}$Se-NMR data on the iron-chalcogenide superconductor K$_{0.8}$Fe$_2$Se$_2$   seems to be consistent with FM fluctuations in the high $T$ paramagnetic phase.\cite{Kotegawa2011} 
    These results suggest that strong FM correlations and fluctuations may be important to detemining $T_{\rm c}$ in the iron-based superconductors. 
    Due to the partial cancellation of the influences of FM and AFM fluctuations in NMR measurements, polarized inelastic neutron scattering measurements are needed to definitively measure the relative strengths of FM and AFM fluctuations in SrCo$_2$As$_2$ and other iron-based superconductors.

   We thank Alan Goldman and Andreas Kreyssig for helpful discussions. The research was supported by the U.S. Department of Energy, Office of Basic Energy Sciences, Division of Materials Sciences and Engineering. Ames Laboratory is operated for the U.S. Department of Energy by Iowa State University under Contract No.~DE-AC02-07CH11358. V. O. thanks the Ames Laboratory and US DOE for providing him  the opportunity to be a visiting scientist at the Ames Laboratory and also thanks the Russian Foundation for Basic Research (No. 15-02-02000) for support.


\begin{thebibliography}{99}
\bibitem{Johnston2010} D. C. Johnston, Adv.  Phys. {\bf 59}, 803 (2010).
\bibitem{Ni2008} N. Ni, S. Nandi, A. Kreyssig, A.~I. Goldman, E.~D. Mun, S. ~L.  Bud'ko, and P.~C. Canfield, Phys. Rev. B {\bf 78}, 014523 (2008).
\bibitem{Rotter2008} M.~Rotter, M.~Tegel, and D.~Johrendt, Phys. Rev. Lett. {\bf101} 107006 (2008).
\bibitem{Sefat2008}A.~S. Sefat, R. Jin, M.~A. McGuire, B. C. Sales, D.~J. Singh, and D. Mandrus, Phys. Rev. Lett. {\bf 101}, 117004 (2008).
\bibitem{Canfield2010} P. ~C. Canfield and S.~L. Bud'ko, Annu. Rev. Condens. Matter Phys. {\bf 1}, 27 (2010).
\bibitem{Stewart2011} G.~R. Stewart, Rev. Mod. Phys. {\bf 83}, 1589 (2011).
\bibitem{Ni2008_2} N. Ni, M. E. Tillman, J. Q. Yan, A. Kracher, S.~T. Hannahs, S.~L. Bud'ko, and P.~C. Canfield, Phys. Rev. B {\bf 78}, 214515 (2008)
\bibitem{Jasper2008}A. Leithe-Jasper, W. Schnelle, C. Geibel, and H. Rosner, Phys. Rev. Lett. {\bf 101}, 207004 (2008).
\bibitem{Hu2011} R. Hu, S. L. Bud'ko, W. E. Straszheim, and P. C. Canfield, Phys. Rev. B {\bf 83}, 094520 (2011).
\bibitem{Krellner2008} C. Krellner, N. Caroca-Canales, A. Jesche, H. Rosner, A. Ormeci, and C. Geibel, Phys. Rev. B {\bf 78}, 100504(R) (2008).
\bibitem{Zhao2008}  J. Zhao, D.-X. Yao, S. Li, T. Hong, Y. Chen, S. Chang, W. Ratcliff, II, J.~W. Lynn, H.~A. Mook, G.~F. Chen, J.~L. Luo, N. L. Wang, E.~W. Carlson, J. Hu, and P. Dai, Phys. Rev. Lett. {\bf 101}, 167203 (2008). 
\bibitem{pandey} A. Pandey, D.~G.~Quirinale, W. Jayasekara, A. Sapkota, M.~G.~Kim, R.~S.~Dhaka, Y. Lee, T.~W.~Heitmann, P.~W.~Stephens, V. Ogloblichev, 
A. Kreyssig, R.~J.~McQueeney, A.~I.~Goldman, A. Kaminski, B.~N.~Harmon, Y. Furukawa, and D.~C.~Johnston, Phys. Rev. B {\bf 88}, 014526 (2013).
\bibitem{INS} W. Jayasekara, Y. Lee, A. Pandey, G. S. Tucker, A. Sapkota, J. Lamsal, S. Calder, D. L. Abernathy, J. L. Niedziela, 
B. N. Harmon, A. Kreyssig, D. Vaknin, D.~C.~Johnston, A.~I.~Goldman and R.~J.~McQueeney, Phys. Rev. Lett. {\bf 111}, 157001 (2013).
\bibitem{Hardy2013} F. Hardy, A. E. B{\"o}hmer, D. Aoki, P. Burger, T. Wolf, P. Schweiss, R. Heid, P. Adelmann, Y. X. Yao, G. Kotliar, J. Schmalian and C. Meingast, Phys. Rev. Lett.  {\bf 111}, 027002 (2013).
\bibitem{Hirano2012}M. Hirano, Y. Yamada, T. Saito, R. Nagashima, T. Konishi, T. Toriyama, Y. Ohta, H. Fukazawa, Y. Kohori, Y. Furukawa, K. Kihou, C-H Lee, A. Iyo and H. Eisaki, J. Phys. Soc. Jpn. {\bf 81}, 054704  (2012). 
\bibitem{T1} $1/T_1$ at each $T$ was determined by fitting the nuclear magnetization $M$ versus time $t$ using the exponential functions $1-M(t)/M(\infty) = 0.1e^{-t/T_1}+0.9e^{-6t/T_1}$ for $^{75}$As and $1-M(t)/M(\infty) = 0.012e^{-t/T_1}+0.068e^{-6t/T_1}+0.206e^{-15t/T_1}+0.714e^{-28t/T_1}$ for $^{59}$Co, where $M(t)$ is the nuclear magnetization at time $t$ after saturation and $M(\infty)$ is the equilibrium nuclear magnetization at $t\to\infty$. 
\bibitem{Singh2008} D. J. Singh and M. H. Du, Phys. Rev. Lett. {\bf 100}, 237003 (2008).
\bibitem{Mazin2008} I.~I.~Mazin, D.~J.~Singh, M.~D.~Johannes, and M.~H.~Du, Phys. Rev. Lett. {\bf 101}, 057003 (2008).
\bibitem{Dong2008} J. Dong, H.~J.~Zhang, G. Xu, Z. Li, G. Li, W. Z. Hu, D. Wu, G. F. Chen, X. Dai, J. L. Luo, Z. Fang, and N. L. Wang, Europhys. Lett. {\bf  83}, 27006 (2008).
\bibitem{Yaresko2009} A.~N.~Yaresko, G.-Q.~Liu, V.~N.~Antonov, and O.~K.~Andersen, Phys. Rev. B {\bf 79}, 144421 (2009).
\bibitem{Neupane2011}M. Neupane, P. Richard, Y.-M. Xu, K. Nakayama, T. Sato, T. Takahashi, A. V. Federov, G. Xu, X. Dai, Z. Fang, Z. Wang, G.-F. Chen, N.-L. Wang, H.-H. Wen, and H. Ding, Phys. Rev. B {\bf 83}, 094522 (2011).
 \bibitem{Ning2009} F. L. Ning, K. Ahilan, T. Imai, A. S. Sefat, R. Jin, M. A. McGuire, B. C. Sales, and D. Mandrus, Phys. Rev. B {\bf 79}, 140506(R)  (2009). 
\bibitem{Tsuda1968} T. Tsuda, A. Hirai, and H. Abe, Phys. Lett. {\bf 26A}, 463 (1968).
\bibitem{Fukai1996} T. Fukai, Y. Furukawa, S. Wada, and K. Miyatani, J. Phys. Soc. Jpn. {\bf 65}, 4067 (1996).
\bibitem{Roy2013} B. Roy, Abhishek Pandey, Q. Zhang, T. W. Heitmann, D. Vaknin,  D. C. Johnston, and Y. Furukawa, Phys. Rev. B {\bf 88}, 174415  (2013). 
\bibitem{Moriya1963} T. Moriya, J. Phys. Soc. Jpn. {\bf 18}, 232 (1963). 
\bibitem{Narath1968}A. Narath and H. T. Weaver, Phys. Rev. {\bf 175}, 378 (1968).
\bibitem{Moriya1963_2} T. Moriya, J. Phys. Soc. Jpn. {\bf 18}, 516 (1963).
\bibitem{baco2as2yuji} V.~K.~Anand, D.~G.~Quirinale, Y. Lee, B.~N.~Harmon, Y. Furukawa, V.~V. Ogloblichev, A. Huq, D.~L.~Abernathy, P.~W.~Stephens, R.~J.~McQueeney, A.~Kreyssig, A.~I.~Goldman, and D.~C.~Johnston, Phys. Rev. B {\bf 90}, 064517 (2014). 
\bibitem{baco2as2imai}  K. Ahilan, T. Imai, A. S. Sefat, and F. L. Ning, Phys. Rev. B {\bf 90}, 014520 (2014).
\bibitem{Jeglic2010}  P. Jegli${\rm \breve{c}}$, A. Poto${\rm \breve{c}}$nik, M. Klanj${\rm \breve{s}}$ek, M. Bobnar, M. Jagodi${\rm \breve{c}}$, K. Koch, H. Rosner, S. Margadonna, B. Lv, A. M. Guloy, and D. Ar${\rm \breve{c}}$on, Phys. Rev. B {\bf 81}, 140511(R) (2010).
\bibitem{Kotegawa2011} H. Kotegawa, Y. Hara, H. Nohara, H. Tou, Y. Mizuguchi, H. Takeya, and Y. Takano,  J. Phys.  Soc. Jpn. {\bf 80}, 043708 (2011).
\end{thebibliography}
\end{document}